\documentclass[12pt]{article}

\usepackage{natbib}
\usepackage{amsmath}
\usepackage{amsfonts}
\usepackage{amssymb}
\usepackage{latexsym}
\usepackage{a4}
\usepackage{ulem}

\usepackage{enumitem}
\setitemize{leftmargin=*}

\def\P{{\text P}}
\def\T{{\text T}}

\def \D{{\mathrm{D}}}

\def \LL{{\cal{L}}}

\def \RR{{\cal{R}}}

\def \UU{\boldsymbol{U}}

\def \Bbeta{{\boldsymbol{\beta}}}
\def \Balpha{{\boldsymbol{\alpha}}}

\def \BI{{\boldsymbol{I}}}
\def \Bf{{\boldsymbol{f}}}
\def \BR{{\boldsymbol{R}}}
\def \Br{{\boldsymbol{r}}}
\def \Bx{{\boldsymbol{x}}}
\def \Bv{{\boldsymbol{v}}}
\def \Bu{{\boldsymbol{u}}}

\begin{document}


\title{{\bf On the equations of motion of dislocations in quasicrystals}}
\author{Eleni Agiasofitou~\footnote{Corresponding author.
{\it E-mail address:}~agiasofitou@mechanik.tu-darmstadt.de.}
\ and
Markus Lazar~\footnote{
{\it E-mail address:}~lazar@fkp.tu-darmstadt.de.}
\\
\\
Heisenberg Research Group,\\ Department of Physics,\\ 
Darmstadt University of Technology, \\
Hochschulstr. 6, D-64289 Darmstadt, Germany
}

\date{ }

\maketitle

\begin{abstract}
In this work we investigate the theory of dynamics of dislocations in quasicrystals.
We consider three models: the elastodynamic model of wave type, the elasto-hydrodynamic model, and the elastodynamic model of wave-telegraph type. Similarities and differences between the three models are pointed out and discussed. Using the framework of linear incompatible elastodynamic theory, the equations of motion of dislocations are deduced for these three models. Especially, the  equations of motion for the phonon and phason elastic distortion tensors and elastic velocity vectors are derived, where the source fields are given in terms of the phonon and phason dislocation density and dislocation current tensors in analogy to the classical theory of elastodynamics of dislocations. The equations of motion for the displacement fields are also obtained.\\


\end{abstract}

\section{Introduction}

Quasicrystals, discovered by \citet{Shechtman1984}, are becoming nowadays an important interdisciplinary research field. Their complicated, fascinating structure, and particular properties have attracted the interest of scientists from several different research areas such as solid state physics, applied mathematics, material science, crystallography, and chemistry. An interesting review on the theory of quasicrystals and its applications is given in the book by \citet{Fan Book}.
\par
The description of the dynamical behavior of quasicrystals challenges the scientific community from their discovery up to now, since there is not a unique opinion about the equations of motion. Based on the theory of hydrodynamics, \citet{Lubensky85} proposed {\it the hydrodynamic model} where phasons play a different role than phonons, since phasons are insensitive to spatial translations.  Experimental studies on phason dynamics based on hydrodynamics have been accomplished by \citet{Francoual et al. 2003, Francoual et al. 2006, deBoissieu2008, deBoissieu et al. 2008}. Hydrodynamics deals with excitations in the long wavelength limit and as a result of the hydrodynamic theory phason modes are always diffusive modes, provided the wavelength is large enough (see, e.g., \citet{deBoissieu2008, deBoissieu et al. 2008}). \citet{Bak 1985a, Bak 1985b} has proposed that phonons and phasons are acoustic modes representing a total of six continuous symmetries (for icosahedral quasicrystals). He has shown that such quasicrystals can always be described as three-dimensional cuts in six-dimensional regular crystals. Based on Bak's arguments, \citet{Ding1993} generalized first the classical theory of compatible linear elastodynamics towards quasicrystals and they proposed the
{\it elastodynamic model} where both, phonons and phasons represent wave propagations. For reasons of clarity, it is necessary as we will see below to specify the elastodynamic model as {\it elastodynamic model of wave type}. On the other hand, \citet{Rochal2002} proposed
the minimal model of the phonon-phason elastodynamics according to, the equations of motion for the phonons are of
wave type while for the phasons are of diffusion type. Later,
\citet{Fan2009} adopted this model under the name
elasto-hydrodynamic model of quasicrystals and it has been used by many
researchers (e.g., \citet{Li2011, Fanetal2012}).
\par
In \citet{AgiasofitouLazar2013}, we have
proposed and described in detail the so-called {\it
elastodynamic model of wave-telegraph type}. According to this
model, phonons are represented by undamped waves, and phasons by waves damped in time and propagating with finite velocity. Therefore, the equations of motion for the phonons are of wave type
and for the phasons of telegraph type. This model possesses  many advantages, among the others, we mention that the influence of the damping in the dynamic behavior of phasons is expressed by the tensor of phason friction coefficients giving the possibility to capture anisotropic effects of the damped waves. Moreover, the characteristic time of damping can be calculated, since it is defined in terms of the phason friction coefficient and the average mass density of the material. This model provides a theory valid in the whole regime of possible wavelengths for the phasons and in addition it overcomes the paradox of the infinite velocity propagation which is implied by the diffusion equation. We should notice that the elastodynamic model of wave type and the elasto-hydrodynamic model can be recovered as asymptotic cases of the elastodynamic model of wave-telegraph type by letting the phason friction coefficient to obtain extreme values.
\par
Dislocations were observed rather early in quasicrystals and the first direct evidence for dislocation motion was reported by \citet{Wollgarten1995} by {\it in-situ} straining experiments using transmission electron microscopy. An overview on the physical properties of quasicrystals and the elasticity theory of dislocations is presented by \citet{Hu2000}. A review in characteristic features of quasicrystalline dislocations and the plastic properties of quasicrystals can be found in \citet{Edagawa2001, Edagawa2007}. Furthermore, an interesting discussion about the complexity of dislocation motion, like the dominant mode responsible for the plastic deformation and the different viewpoints on this aspect, is given by \citet{Guyot2003} and the references therein. Simulations on dislocation motion closely related to crack propagation can be found in \citet{Mikulla et al. 1998a, Mikulla et al. 1998b}. A review on experimental results on dislocation dynamics (in icosahedral Al-Pd-Mn single quasicrystals) was reported by \citet{Messerschmidt et al. 2001}.
 \par
 It is a fact that, at the microscopic scale, plasticity occurs by dislocation motion. Therefore, the plastic fields should also be considered in a consistent theory of studying dislocations in quasicrystals. For defects, the plastic distortion tensors give the so-called eigenstrains \citep{Mura}. Using a generalized eigenstrain method towards quasicrystals, \citet{Ding et al 1995} (see also \citet{Ding et al 1994, Hu2000}) calculated the displacement fields of (static) straight dislocations in quasicrystals induced by eigenstrains (plastic distortions). \cite{Agiasofitou} have developed the generalized theory of incompatible elastodynamics of quasicrystals. They have given the basic theoretical framework for dislocation dynamics in quasicrystals, generalizing the classical theory of incompatible elastodynamics \citep{Kossecka1975, Kossecka-deWit1977} towards the theory of quasicrystals. In this framework, taking the plastic fields into account is essential, since a dislocation is the elementary carrier of plasticity. Among others the dynamical Eshelby stress tensor, Peach-Koehler force and J-integral have been derived for the elastodynamic model of wave type and the elasto-hydrodynamic model.
 \par
 The present work is posed in the framework of incompatible elasticity theory of moving dislocations in quasicrystals. Throughout the paper we examine three models: the
elastodynamic model of wave type, the elasto-hydrodynamic model, and the
elastodynamic model of wave-telegraph type. Our main aim is, for these models, to derive the
equations of motion for the elastic fields, where the dislocation density and dislocation current tensors are the source terms in analogy to the classical theory of elastodynamics of dislocations. The elastic fields and the displacement fields caused by moving dislocations are important in the modeling of dislocation-based plasticity theory. All three models are presented and investigated providing in this way a new insight for their consideration and comparison.
\par
This paper
is organized as follows: Section 2 deals with the necessary preliminaries
concerning the incompatible linear elasticity of moving dislocations in quasicrystals. In Section 3 we set the stage for the three models and we deduce the field equations in terms of the momenta and the stresses. The differences in the basic assumptions of the three models are pointed out. In Section 4 we derive the equations of motion for the elastic fields. The type of the obtained partial differential equations and the qualitative (mathematical and physical) differences and similarities are discussed. Section 5 is devoted to the derivation of the equations of motion for the displacement fields. The equations of motion in the case of the compatible elastodynamics of quasicrystals are also given.

\section{Incompatible elasticty of moving dislocations}
In this section, we give the necessary relevant material concerning the
theory of incompatible linear elasticity of moving dislocations
in quasicrystals (see also \cite{Agiasofitou}),
thus making our exposition self-contained.

The displacement field $\UU$ in a quasicrystal is constituted by two components, the usual {\it phonon displacement field} $u_i^\|(\Bx,t) \in E_\|$, which belongs to the physical or parallel space $E_\|$,
and the {\it phason displacement field} $u_i^\bot (\Bx, t) \in E_\bot$ that
belongs to the perpendicular space $E_\bot$, that is
\begin{equation}\label{U}
\UU=(\Bu^\|, \Bu^\bot) \ \in E_\| \oplus E_\bot.
\end{equation}
All quantities depend on the physical space coordinates $\Bx \in E_\|$ and the
time $t$. In the theory of incompatible elasticity of quasicrystals, the total distortion tensors for the
fields of phonon ${\beta}^{\|\, \T}_{ij}$, and phason ${\beta}^{\bot \T}_{ij}$,
are decomposed into {\it the elastic distortion tensors} $\beta^\|_{ij}, \ \beta^\bot_{ij}$,
and {\it the plastic distortion tensors} ${\beta}^{\|\, \P}_{ij}, \ {\beta}^{\bot \P}_{ij}$,
respectively
\begin{alignat}{2}\label{ugrad}
u^\|_{i,j}:={\beta}^{\|\, \T}_{ij}=\beta^\|_{ij}+{\beta}^{\|\, \P}_{ij}, \qquad
u^\bot_{i,j}:={\beta}^{\bot \T}_{ij}=\beta^\bot_{ij}+{\beta}^{\bot \P}_{ij},
\end{alignat}
where comma denotes differentiation with respect to the spatial physical
coordinates. In addition, in incompatible elastodynamics of quasicrystals the differentiation of the displacement fields with respect to the time $t$ will give the total velocities, $v_i^{\|\, \T}$ and $v_i^{\bot\, \T}$, which can be decomposed into {\it the elastic velocities}, $v^\|_i$ and $v^\bot_i$, and {\it the plastic} or { \it initial velocities} \citep{Kossecka1975, Kossecka-deWit1977}, $v_i^{\|\, \P}$ and $v_i^{\bot\, \P}$, for the phonon and phason fields, respectively
\begin{alignat}{2}\label{udot}
\dot{u}_i^\|:=v_i^{\|\, \T}=v_i^\|+v_i^{\|\, \P},\qquad
\dot{u}_i^\bot:=v_i^{\bot\, \T}=v_i^\bot+v_i^{\bot\, \P}.
\end{alignat}
A superimposed dot denotes differentiation with respect to time.
\par
The elastic fields are state quantities in contrast to the plastic fields. The elastic as well as the plastic fields satisfy certain incompatibility conditions \citep{Agiasofitou}.
{\it The incompatibility conditions} for the elastic fields, which are useful for our further calculations are given by
\begin{alignat}{2}
&\alpha^\|_{ij}=e_{jkl}\beta^\|_{il,k},  \qquad     &&\alpha^\bot_{ij}=e_{jkl}\beta^\bot_{il,k},     \label{IC1}\\
&I^\|_{ij}={\dot \beta}^\|_{ij}-v^\|_{i,j},      \qquad  &&I^\bot_{ij}={\dot \beta}^\bot_{ij}-v^\bot_{i,j}.      \label{IC2}
\end{alignat}
The incompatibility conditions for the plastic fields read
\begin{align}\label{IC1-p}
\alpha^\|_{ij}=-e_{jkl}\beta^{\|\, \P}_{il,k},
\quad  \alpha^\bot_{ij}=-e_{jkl}\beta^{\bot\,\P}_{il,k},\qquad
I^\|_{ij}=-{\dot \beta}^{\|\,\P}_{ij}+v^{\|\,\P}_{i,j},
\quad I^\bot_{ij}=-{\dot \beta}^{\bot\,\P}_{ij}+v^{\bot\,\P}_{i,j}\,.
\end{align}
In the above equations, $\alpha^\|_{ij}$ and $\alpha^\bot_{ij}$ are {\it the phonon} and {\it the phason dislocation density tensors}, $I^\|_{ij}$ and $I^\bot_{ij}$  are {\it the phonon} and {\it the phason
dislocation current tensors}, respectively and $e_{jkl}$ is {\it the permutation tensor} or {\it Levi-Civita tensor}. The dislocation density and dislocation current tensors are state
quantities, because they can be measured in the present state of the body.
On the contrary, the plastic distortion tensors and the plastic velocity vectors are
not state quantities, since one has to know the prior history of the body to
measure them.

\par
In addition, the dislocation density and the dislocation current tensors satisfy
{\it the Bianchi identities}, which are geometrical constraints for these quantities
\begin{align}\label{Bianchi2}
\alpha^\|_{ij,j}=0, \quad   \alpha^\bot_{ij,j}=0,\qquad
{\dot \alpha}^\|_{ij}=e_{jkl}I^\|_{il,k},\quad {\dot \alpha}^\bot_{ij}=e_{jkl}I^\bot_{il,k}.
\end{align}
The dislocation density tensors $\Balpha^\|$ and $\Balpha^\bot$, and the
dislocation current tensors $\BI^\|$ and $\BI^\bot$ may describe
straight dislocations, dislocation loops and an arbitrary distribution of
dislocations in quasicrystals. Dislocations can
move uniformly (constant dislocation velocity) and
non-uniformly (non-constant dislocation velocity which consequences dislocation acceleration).
\par

For example, for a non-stationary dislocation loop $L(t)$
the plastic distortion tensor, the plastic velocity vector,
the dislocation density and the dislocation current tensors,
given by~\citet{Kossecka-deWit1977} can be generalized towards quasicrystals.
The corresponding phonon and phason quantities are given by the following
time-dependent surface and line integrals
\begin{alignat}{2}
\label{B-L}
&\beta_{ij}^{\|\, \P}(\Br,t)=-\int_{S(t)}b_i^{\|\,}\delta(\BR)\, d S'_j\,,
&&\beta_{ij}^{\bot\, \P}(\Br,t)=-\int_{S(t)}b_i^{\bot\,}\delta(\BR)\, d S'_j\,,\\
\label{v-L}
&v_{i}^{\|\, \P}(\Br,t)=\int_{S(t)} b_i^{\|\,} V'_j\, \delta(\BR)\, d S'_j\,,
&&v_{i}^{\bot\, \P}(\Br,t)=  \int_{S(t)} b_i^{\bot\,} V'_j\, \delta(\BR)\, d S'_j\,,\\
\label{A-L}
&\alpha_{ij}^{\|}(\Br,t)=\oint_{L(t)} b_i^{\|\,} \delta(\BR)\, d L'_j\,,
&&\alpha_{ij}^{\bot}(\Br,t)=\oint_{L(t)}b_i^{\bot\,}\delta(\BR)\, d L'_j\,,\\
\label{I-L}
&I_{ij}^{\|}(\Br,t)=\oint_{L(t)} e_{jkl}\, b_i^{\|\,} V'_k\,\delta(\BR)\, d L'_l\,,\quad
&&I_{ij}^{\bot}(\Br,t)=\oint_{L(t)} e_{jkl}\, b_i^{\bot\,} V'_k\,\delta(\BR)\, d L'_l\,,
\end{alignat}
where $S(t)$ is the time-dependent surface bounded by $L(t)$, $b_i^\|$ and $b_i^\bot$ are the phonon and phason components of the Burgers vector respectively, $\delta$ is the Dirac-delta function, $\BR=\Br-\Br'(t)$ is the relative vector between the time-dependent point $\Br'(t)$ of
the moving dislocation loop and the position vector $\Br$, $V'_i=\dot{r}'_i(t)$ is the dislocation velocity at any point $\Br'$
of the loop, $d L'_j=d L_j(\Br'(t))$ is the time-dependent line element along the loop
and  $d S'_j=d S_j(\Br'(t))$ is the corresponding time-dependent surface
element. Moreover,
the surface $S$ represents the area swept by the loop $L$ during its
motion and may be called the dislocation surface.
The plastic distortion caused by a dislocation loop is concentrated
at the surface $S$.
The surface $S$ is what determines the history of the plastic distortion
of a dislocation loop.
\section{Field equations}
In this section, we present the three models and we deduce the field equations in terms of the momenta and the stresses in the framework of linear incompatible elasticity theory of moving dislocations in the presence of external forces.
\par
 Let us consider the Lagrangian density $\LL$ as a smooth function of the physical state quantities, that is $\LL=\LL(\Bv^\|, \Bv^\bot, \Bbeta^\|, \Bbeta^\bot).$ The Lagrangian density can be given in terms of the kinetic energy density $T$, the elastic energy density $W$ and the potential $V$ of the external forces
\begin{equation}
\label{L}
\LL=T-W-V.
\end{equation}
The elastic energy density $W$ (for the unlocked state) is considered
as a quadratic function of the phonon and phason strains (\cite{Ding1993, Agiasofitou})
\begin{equation}\label{w}
W=\frac{1}{2}\beta^\|_{ij}C_{ijkl}\beta^\|_{kl}+\beta^\|_{ij}D_{ijkl}\beta^\bot_{kl}+\frac{1}{2}\beta^\bot_{ij}E_{ijkl}\beta^\bot_{kl}
\end{equation}
with the tensors of elastic constants to possess the following symmetries
\begin{align}\label{Coef}
C_{ijkl}=C_{klij}=C_{ijlk}=C_{jikl},\quad D_{ijkl}=D_{jikl}, \quad E_{ijkl}=E_{klij}.
\end{align}
The potential $V$ of the external forces may be introduced as
\begin{equation}\label{V1}
V=-u^\|_i f_i^\|-u_i^\bot f_i^\bot,
\end{equation}
where $f_i^\|$ is {\it the conventional
(phonon) body force density} and $f_i^\bot$ is {\it a generalized (phason) body force density}.\\
The kinetic energy density $T$ admits phonon and phason contributions
\begin{equation}\label{T1}
T=\frac{1}{2}\rho(v^\|_i)^2+\frac{1}{2}\rho(v^\bot_i)^2,
\end{equation}
where $\rho$ is the average mass density of the material. \citet{Rochal2000} proposed that the phasons should possess
a phason mass density, the  effective phason density  $\rho_{\text {eff}}$, which may be different from $\rho$. However, later in their suggested minimal model of phonon-phason elastodynamics \citep{Rochal2002}, they supported that the phason momentum is not conserved, implying $\rho_{\text {eff}}=0$. In general, the meaning of this quantity is not clear and its measurement is difficult as it is mentioned in \cite{Fan Book}.  In the case that the phasons indeed possess a different mass density, $\rho_{\text {eff}}$, the form of the kinetic energy density~(\ref{T1}) as well as of the corresponding derived equations bellow can be easily modified.
\par
The constitutive relations for the considered Lagrangian density are
\begin{align}
&p_i^\|=\frac{\partial T}{\partial v^{\|}_i}=\rho v^\|_i, \label{CR1}\\
&p^\bot_i=\frac{\partial T}{\partial v^\bot_i}=\rho v^\bot_i, \label{CR2}\\
&\sigma^\|_{ij}=\frac{\partial W}{\partial \beta^\|_{ij}}=C_{ijkl}\beta^\|_{kl}+D_{ijkl}\beta^\bot_{kl}, \label{CR3}\\
&\sigma^\bot_{ij}=\frac{\partial W}{\partial \beta^\bot_{ij}}=D_{klij}\beta^\|_{kl}+E_{ijkl}\beta^\bot_{kl} \label{CR4}.
\end{align}
In the above relations, $p_i^\|$ and $p^\bot_i$ are {\it the phonon} and {\it phason momentum vectors},
and $\sigma^\|_{ij}$ and $\sigma^\bot_{ij}$ are {\it the phonon} and {\it phason stress tensors}, respectively. We note that the phonon stress tensor is symmetric,
$\sigma^\|_{ij}=\sigma^\|_{ji}$, while the phason stress tensor is asymmetric, $\sigma^\bot_{ij}\neq\sigma^\bot_{ji}$, (see, e.g., \cite{Ding1993}).\\
The elastic energy density $W$ has the same form for the three models, while the kinetic energy density $T$ as well as the use of the potential $V$ and the consideration of additional phason forces, like phason ``bulk forces"  \citep{Rochal2002} and phason frictional forces \citep{AgiasofitouLazar2013} are different for the considered models. Therefore, it is necessary from now on to examine the three models separately.
\begin{itemize}
\item Elastodynamic model of wave type:\\
For the elastodynamic model of wave type, the eqs.~(\ref{L})--(\ref{CR4}) are valid. The corresponding Euler-Lagrange equations with respect to the displacement fields are
\begin{align}
&&\label{EL1}\frac{\partial \LL}{\partial u^\|_i}-\frac{\partial}{\partial t}
\bigg(\frac{\partial \LL}{\partial \dot{u}^\|_i}\bigg)-
\frac{\partial}{\partial x_j}\bigg(\frac{\partial \LL}{\partial u^\|_{i,j}}\bigg)=0,\\
&&\label{EL2} \frac{\partial \LL}{\partial u^\bot_i}
-\frac{\partial}{\partial t}\bigg(\frac{\partial \LL}{\partial \dot{u}^\bot_i}\bigg)-
\frac{\partial}{\partial x_j}\bigg(\frac{\partial \LL}{\partial u^\bot_{i,j}}\bigg)=0.
\end{align}
Using the relations (\ref{ugrad}) and (\ref{udot}) and the constitutive relations~(\ref{CR1})--(\ref{CR4}),
the Euler-Lagrange equations~(\ref{EL1}) and (\ref{EL2})
give {\it the field equations for the elastodynamics of wave type of quasicrystals} in terms of the momenta and the stresses
\begin{align}\label{M1}
&\dot{p}_i^\|-\sigma^\|_{ij,j}=f_i^\|,\\
\label{M2}
&\dot{p}^\bot_i-\sigma^\bot_{ij,j}=f_i^\bot.
\end{align}
The above equations can explicitly be written in terms of the elastic fields
$\Bv^\|, \Bv^\bot, \Bbeta^\|$ and $\Bbeta^\bot$ as follows
\begin{align}\label{EqM1}
& \rho \dot{v}^\|_i-C_{ijkl}\beta^\|_{kl,j}-D_{ijkl}\beta^\bot_{kl,j}=f_i^\|, \\
\label{EqM2}
& \rho \dot{v}^\bot_i-D_{klij}\beta^\|_{kl,j}-E_{ijkl}\beta^\bot_{kl,j}=f_i^\bot.
\end{align}
\item Elasto-hydrodynamic model:\\
According to this model \citep{Rochal2002}, only the phonon field contributes to the kinetic energy density
\begin{equation}\label{T2}
T=\frac{1}{2}\rho(v^\|_i)^2,
\end{equation}
since the phason momentum is zero, $p_i^\bot=0$. In addition, the attenuation of the phason modes is described by the introduction of a so-called {\it phason ``bulk force"} \cite{Rochal2002}
\begin{align}\label{force}
f_i^\bot=-\D {v}_i^\bot,
\end{align}
where $\D$ is the {\it phason friction coefficient}.
\par
Eq.~(\ref{M1}) remains and eq.~(\ref{M2}) holds now for $\dot{p}_i^\bot=0$ and $f_i^\bot$ as it is given by eq.~(\ref{force}). The corresponding {\it field equations for the elasto-hydrodynamics of quasicrystals} in terms of the momentum (phonon) and the stresses are
\begin{align}
\label{M3}
 \dot{p}_i^\|-\sigma^\|_{ij,j}= f_i^\|,\\
\label{M4}
\sigma^\bot_{ij,j}=\D {v}_i^\bot.
\end{align}
The constitutive relations~(\ref{CR1}), (\ref{CR3}), and (\ref{CR4}) are still valid. Hence, the above equations using the explicit form of eqs.~(\ref{CR1}), (\ref{CR3}) and (\ref{CR4}), become
\begin{align}\label{EqM3}
 \rho \dot{v}^\|_i-C_{ijkl}\beta^\|_{kl,j}-D_{ijkl}\beta^\bot_{kl,j}&= f_i^\|, \\
\label{EqM4}
\D {v}_i^\bot-
 D_{klij}\beta^\|_{kl,j}-E_{ijkl}\beta^\bot_{kl,j}&=0.
\end{align}
{\bf Remark:} One can observe in eq.~(\ref{force}) that a (phason) body force depends on the velocity (internal field). Therefore,  $f_i^\bot$ is in fact an internal force but it is used in the place of an external force and inserted ``by hand" in the derived equations of motion. Eqs.~(\ref{M3}) and (\ref{M4}) have actually been derived from eqs.~(\ref{M1}) and (\ref{M2}) with the substitutions that we have mentioned above, and not directly from Euler-Lagrange equations.  At this point there is a contradiction from the physical point of view; that a force depending on the velocity (internal field) is considered as an external force. Moreover, such an identification is not allowed in the framework of calculus of variations.
\item Elastodynamic model of wave-telegraph type:\\
The main characteristics of this model \citep{AgiasofitouLazar2013} are that the kinetic energy density consists of phonon and phason contributions as it is given by eq.~(\ref{T1}) and dissipative processes for the phason fields are taken additionally into account. The consideration of the dissipation consequences the appearance of {\it phason frictional forces}. These are linear functions of the phason velocities
\begin{align}\label{fr}
f_i^{\text{fr}}=-\D_{ij}\,v^\bot_j,
\end{align}
where $\D_{ij}$ is the {\it tensor of phason friction coefficients} which is symmetric (see, e.g., \cite{Landau Statistical Physics}), that is $\D_{ij}=\D_{ji}$. The phason frictional forces can be written as the derivative of a quadratic function, namely of the so-called {\it dissipative function density} $\RR=\frac{1}{2}\D_{ij}\,v^\bot_i v^\bot_j$, \citep{Landau Statistical Physics, Landau Mechanics, Landau Elasticity, Goldstein} with respect to the corresponding velocities. That is,
$f_i^{\text{fr}}=-{\partial \RR}/{\partial v^\bot_i}.$
\\
Eqs.~(\ref{L})--(\ref{CR4}) are valid also for this model. In this case, the Euler-Lagrange equations have to be modified, since dissipative processes for the phason fields are considered, as follows
\begin{align}
&\label{EL3}\frac{\partial \LL}{\partial u^\|_i}-\frac{\partial}{\partial t}
\bigg(\frac{\partial \LL}{\partial \dot{u}^\|_i}\bigg)-
\frac{\partial}{\partial x_j}\bigg(\frac{\partial \LL}{\partial u^\|_{i,j}}\bigg)=0,\\
&\label{EL4} \frac{\partial \LL}{\partial u^\bot_i}
-\frac{\partial}{\partial t}\bigg(\frac{\partial \LL}{\partial \dot{u}^\bot_i}\bigg)-
\frac{\partial}{\partial x_j}\bigg(\frac{\partial \LL}{\partial u^\bot_{i,j}}\bigg)=\frac{\partial \RR}{\partial  \dot{u}^\bot_i}.
\end{align}
The Euler-Lagrange equation for the phason fields (eq.~(\ref{EL4})) differs from the usual form of the Euler-Lagrange equations by the presence of the derivative of the dissipative function density on the right.
\par
Using the relations (\ref{ugrad}) and (\ref{udot}) and the constitutive relations~(\ref{CR1})--(\ref{CR4}),
the Euler-Lagrange equations~(\ref{EL3}) and (\ref{EL4}) give {\it the  field equations for the elastodynamics of wave-telegraph type of quasicrystals} in terms of the momenta and the stresses
\begin{align}
\label{M5}
\dot{p}_i^\|-\sigma^\|_{ij,j}=f^\|_i,\\
\label{M6}
\dot{p}_i^\bot-\sigma^\bot_{ij,j}-f_i^{\text{fr}}=f_i^\bot.
\end{align}
Eqs.~(\ref{M5}) and (\ref{M6}) can further be written in terms
of the elastic fields, using the explicit forms of the constitutive relations~(\ref{CR1})--(\ref{CR4}),
\begin{align}
\label{EqM5}
\rho \dot{v}^\|_i-C_{ijkl}\beta^\|_{kl,j}-D_{ijkl}\beta^\bot_{kl,j}&=f^\|_i, \\
\label{EqM6}
\rho \dot{v}^\bot_i+\D_{ij} {v}_j^\bot-
D_{klij}\beta^\|_{kl,j}-E_{ijkl}\beta^\bot_{kl,j}&=f_i^\bot.
\end{align}
{\bf Remark:} We provide here some qualitative remarks concerning the mathematical formulation of this model. The appearance of phason frictional forces (dissipative forces) leads to the damping of the phason waves, as we will also see in the next section. The anisotropic tensor $\D_{ij}$ accounts for anisotropic effects of the damped waves (see also \cite{AgiasofitouLazar2013}). In the isotropic case, $\D_{ij}$ reads: $\D_{ij}=\D\delta_{ij}$, $\D>0$. $\D=1/\Gamma_w$, where $\Gamma_w$ is the so-called kinetic coefficient of the phason field of the material defined by \citet{Lubensky85} (see \cite{AgiasofitouLazar2013}).  The field equations are derived from a variational principle with dissipation. In this case, attention should be paid to the Euler-Lagrange equations, which are different from the usual ones, because dissipative processes for the phasons fields are also taken into account.
\end{itemize}

\section{The equations of motion for the elastic fields}
This section is devoted to the derivation of the equations of motion for the elastic fields of dislocations
in quasicrystals in the presence of external forces. In the derived equations, the source terms for
the elastic fields are given in terms of the dislocation density
and dislocation current tensors
in analogy to the classical theory of elastodynamics of dislocations
(see, e.g., \citet{Mura, Kossecka77b, XM83,XM90,Lazar2011, Lazar2012}). Thus, the phonon and phason dislocation density and dislocation current tensors will be the ``input" in the equations of motion of the elastic fields of dislocations. For dislocation loops, they are given by Eqs.~(\ref{A-L}) and (\ref{I-L}). It should be emphasized that the equations of motion, derived in this section, are a straightforward generalization of the equations of motion of the elastic fields of the classical elastodynamics of dislocations towards the theory of quasicrystals.
\par
We start with some particular auxiliary identities, stemming from
the incompatibility conditions~(\ref{IC1}) and (\ref{IC2}), which are necessary
for the mathematical procedure that follows. One of those identities can be obtained
if we differentiate eqs.~(\ref{IC2}) with respect to time $t$
\begin{align} \label{Eq104 and 111}
\dot{v}^\|_{i,m}={\ddot \beta}^\|_{im}-\dot{I}^\|_{im},   \qquad  \dot{v}^\bot_{i,m}={\ddot \beta}^\bot_{im}-\dot{I}^\bot_{im}.
\end{align}
If we multiply eq.~(\ref{IC1}) by the permutation tensor $e_{mjk}$,
we obtain the following relations
\begin{align}\label{Eq100}
e_{mjk}\alpha^\|_{im}=\beta^\|_{ik,j}-\beta^\|_{ij,k}, \qquad e_{mjk}\alpha^\bot_{im}=\beta^\bot_{ik,j}-\beta^\bot_{ij,k},
\end{align}
which can give the desired auxiliary equations, after a differentiation with respect to $x_j$,
\begin{align}\label{Eq126}
\beta^\|_{kl,mj}=\beta^\|_{km,lj}+e_{pml}\alpha^\|_{kp,j}, \qquad  \beta^\bot_{kl,mj}=\beta^\bot_{km,lj}+e_{pml}\alpha^\bot_{kp,j}.
\end{align}
After a differentiation with respect to $x_j$, the incompatibility conditions~(\ref{IC2}) give the following useful relations
\begin{align}\label{Eq115 and 116}
{\dot \beta}^\|_{kl,j}=I^\|_{kl,j}+v^\|_{k,lj},      \quad  {\dot \beta}^\bot_{kl,j}=I^\bot_{kl,j}+v^\bot_{k,lj}.
\end{align}
\begin{itemize}
\item Elastodynamic model of wave type:\\
If we differentiate eqs.~(\ref{EqM1}) and (\ref{EqM2}) with respect to $x_m$, we obtain
\begin{align}\label{Eq103}
& \rho \dot{v}^\|_{i,m}-C_{ijkl}\beta^\|_{kl,jm}-D_{ijkl}\beta^\bot_{kl,jm}=f_{i,m}^\|, \\
\label{Eq110}
& \rho \dot{v}^\bot_{i,m}-D_{klij}\beta^\|_{kl,jm}-E_{ijkl}\beta^\bot_{kl,jm}=f_{i,m}^\bot.
\end{align}
Substituting eqs.~(\ref{Eq104 and 111}) and (\ref{Eq126}) into eqs.~(\ref{Eq103}) and (\ref{Eq110}),
we obtain the following equations
\begin{align}
\label{Eq127}
&\rho \ddot{\beta}_{im}^\|-C_{ijkl}\beta_{km,lj}^\|-D_{ijkl}\beta_{km,lj}^\bot=
e_{pml}\large(C_{ijkl}{\alpha}^\|_{kp,j}+D_{ijkl}{\alpha}^\bot_{kp,j}\large)+\rho \dot{I}^\|_{im}+f_{i,m}^\|,\\
\label{Eq129}
&\rho \ddot{\beta}_{im}^\bot-E_{ijkl}\beta_{km,lj}^\bot-D_{klij}\beta_{km,lj}^\|=
e_{pml}\large(E_{ijkl}{\alpha}^\bot_{kp,j}+D_{klij}{\alpha}^\|_{kp,j}\large)+ \rho \dot{I}^\bot_{im}+f_{i,m}^\bot .
\end{align}
{\it Eqs.~(\ref{Eq127}) and ~(\ref{Eq129}) are the equations of motion for the elastic distortion tensors
$\Bbeta^\|$ and $\Bbeta^\bot$ for the elastodynamics of wave type of quasicrystals}. It can be seen that the sources for the elastic distortion tensors are the gradient of the dislocation density tensors
$\Balpha^\|$ and $\Balpha^\bot$, the time derivative of the dislocation current tensors
$\BI^\|$ and $\BI^\bot$ and if external forces are present, the gradient of the
forces $\Bf^\|$ and $\Bf^\bot$.

On the other hand, if we differentiate eqs.~(\ref{EqM1}) and (\ref{EqM2}) with respect to time $t$ and we use eqs.~(\ref{Eq115 and 116}), we obtain
\begin{align}
\label{Eq117}
&\rho \ddot{v}_i^\|-C_{ijkl}v_{k,lj}^\|-D_{ijkl}v_{k,lj}^\bot=
C_{ijkl}I^\|_{kl,j}+D_{ijkl}I^\bot_{kl,j}+\dot{f}_i^\|,\\
\label{Eq118}
&\rho \ddot{v}_i^\bot-E_{ijkl}v_{k,lj}^\bot-D_{klij}v_{k,lj}^\|=
D_{klij}I^\|_{kl,j}+E_{ijkl}I^\bot_{kl,j}+\dot{f}_i^\bot.
\end{align}
{\it Eqs.~(\ref{Eq117}) and (\ref{Eq118}) are the equations of motion for the elastic velocity vectors $\Bv^\|$ and $\Bv^\bot$ for the elastodynamics of wave type of quasicrystals}. The sources for the elastic velocity vectors are the gradient of the dislocation current tensors
$\BI^\|$ and $\BI^\bot$ and if external forces exist, the time derivative of the
forces $\Bf^\|$ and $\Bf^\bot$.
\par
The two systems of eqs.~(\ref{Eq127}) and (\ref{Eq129}), and eqs.~(\ref{Eq117}) and (\ref{Eq118}) are coupled
partial differential equations of wave type. The coupling of the phonon and phason fields is achieved through the constitutive tensor $D_{ijkl}$. The mathematical character of equations~(\ref{Eq127})--(\ref{Eq118}) is hyperbolic and they describe elastodynamic waves. It is well-known that elastodynamic waves propagate with finite velocities. There always exists a time-delay in order for a change in the elastodynamic conditions initiated at a point of the space produces an effect at any other point of the space. This time-delay is called {\it elastodynamic retardation} ~(see, e.g., \cite{Lazar2012}).

\item Elasto-hydrodynamic model:\\
We follow an analogous procedure as previously in order to derive the corresponding equations of motion based on the elasto-hydrodynamic model.  Using eqs.~(\ref{EqM3}) and (\ref{EqM4}), and  eqs.~(\ref{Eq104 and 111}) and (\ref{Eq126}), we obtain {\it the equations of motion for the elastic distortion tensors
$\Bbeta^\|$ and $\Bbeta^\bot$ for the elasto-hydrodynamics of quasicrystals}
\begin{align}
\label{Eq130}
&\rho \ddot{\beta}_{im}^\|-C_{ijkl}\beta_{km,lj}^\|-D_{ijkl}\beta_{km,lj}^\bot=
e_{pml}\large(C_{ijkl}{\alpha}^\|_{kp,j}+D_{ijkl}{\alpha}^\bot_{kp,j}\large)+\rho \dot{I}^\|_{im}+f_{i,m}^\|,\\
\label{Eq131}
&\D\dot{\beta}_{im}^\bot-E_{ijkl}\beta_{km,lj}^\bot-D_{klij}\beta_{km,lj}^\|=
e_{pml}\large(E_{ijkl}{\alpha}^\bot_{kp,j}+D_{klij}{\alpha}^\|_{kp,j}\large)+ \D I^\bot_{im}.
\end{align}
The sources for the elastic distortion tensors are the gradient of the dislocation density tensors
$\Balpha^\|$ and $\Balpha^\bot$, the time derivative of the phonon dislocation current tensor
$\BI^\|$, the phason dislocation current tensor itself $\BI^\bot$ and the gradient of the external (phonon) force $\Bf^\|$. \\
On the other hand, using eqs.~(\ref{EqM3}) and (\ref{EqM4}), and  eq.~(\ref{Eq115 and 116}), we get {\it the equations of motion for the elastic velocity vectors $\Bv^\|$ and $\Bv^\bot$ for the elasto-hydrodynamics of quasicrystals}
\begin{align}
\label{Eq123}
&\rho \ddot{v}_i^\|-C_{ijkl}v_{k,lj}^\|-D_{ijkl}v_{k,lj}^\bot=
C_{ijkl}I^\|_{kl,j}+D_{ijkl}I^\bot_{kl,j}+\dot{f}_i^\|,\\
\label{Eq124}
&\D \dot{v}_i^\bot-E_{ijkl}v_{k,lj}^\bot-D_{klij}v_{k,lj}^\|=
D_{klij}I^\|_{kl,j}+E_{ijkl}I^\bot_{kl,j}.
\end{align}
The sources for the elastic velocity vectors are the gradient of the dislocation current tensors
$\BI^\|$ and $\BI^\bot$ and the time derivative of the external (phonon)
force $\Bf^\|$.
It is obvious that the phason external force is absent from the sources in eqs.~(\ref{Eq131}) and (\ref{Eq124}), since it has been already used through the relation~(\ref{force}).

Furthermore, we see that the equations of motion for the elastic phonon fields $\Bbeta^\|$ and  $\Bv^\|$,
eqs.~(\ref{Eq130}) and (\ref{Eq123}),
are partial differential equations of wave type.
While the equations of motion for the elastic phason fields
$\Bbeta^\bot$ and  $\Bv^\bot$,
eqs.~(\ref{Eq131}) and (\ref{Eq124}),
are partial differential equations of diffusion type.
The equations of motion for the phonon and phason fields are coupled
through the constitutive tensor $D_{ijkl}$. It should be pointed out that eqs.~(\ref{Eq130})--(\ref{Eq124}) have a different mathematical
character. Eqs.~(\ref{Eq130}) and (\ref{Eq123}) are hyperbolic, while eqs.~(\ref{Eq131}) and (\ref{Eq124}) are parabolic.
The diffusive type equations~(\ref{Eq131}) and (\ref{Eq124})
have an important disadvantage that the phason fields propagate
with infinite velocity (see, e.g.,~\cite{Strauss}).  This is unphysical
and gives no retardation and causality, as it should happen in a realistic physical model.
Thus, a diffusion type equation cannot be a realistic approximation
for the dynamics of phason fields.

\item Elastodynamic model of wave-telegraph type:\\
We use an analogous procedure with the previous models
in order to derive the equations of motion for the elastic fields for the  considered model. Using eqs.~(\ref{EqM5}) and (\ref{EqM6}),
and eqs.~(\ref{Eq104 and 111}) and (\ref{Eq126}), we obtain {\it the equations of motion for the elastic distortion tensors
$\Bbeta^\|$ and $\Bbeta^\bot$ for the elastodynamics of wave-telegraph type of quasicrystals}
\begin{align}
\label{B-H-pho}
&\rho \ddot{\beta}_{im}^\|-C_{ijkl}\beta_{km,lj}^\|-D_{ijkl}\beta_{km,lj}^\bot=
e_{pml}\large(C_{ijkl}{\alpha}^\|_{kp,j}+D_{ijkl}{\alpha}^\bot_{kp,j}\large)+\rho \dot{I}^\|_{im}+f^\|_{i,m},\\
\label{B-H-pha}
&\rho \ddot{\beta}_{im}^\bot+\D_{ij}\dot{\beta}_{jm}^\bot-E_{ijkl}\beta_{km,lj}^\bot-D_{klij}\beta_{km,lj}^\|=
e_{pml}\large(E_{ijkl}{\alpha}^\bot_{kp,j}+D_{klij}{\alpha}^\|_{kp,j}\large)
+ \rho \dot{I}^\bot_{im}+ \D_{ij} I^\bot_{jm}+f^\bot_{i,m}.
\end{align}
The gradient of the dislocation density tensors $\Balpha^\|$ and $\Balpha^\bot$,
the time derivative of the dislocation current tensors $\BI^\|$ and $\BI^\bot$, the phason dislocation current tensor itself $\BI^\bot$, and the gradient of the forces $\Bf^\|$ and $\Bf^\bot$ enter as sources in eqs.~(\ref{B-H-pho}) and (\ref{B-H-pha}).\\
On the other hand, using eqs.~(\ref{EqM5}) and (\ref{EqM6}), and eq.~(\ref{Eq115 and 116}), we get {\it the equations of motion for the elastic velocity vectors $\Bv^\|$ and $\Bv^\bot$ for the elastodynamics of wave-telegraph type of quasicrystals}
\begin{align}
\label{v-H-pho}
&\rho \ddot{v}_i^\|-C_{ijkl}v_{k,lj}^\|-D_{ijkl}v_{k,lj}^\bot=
C_{ijkl}I^\|_{kl,j}+D_{ijkl}I^\bot_{kl,j}+\dot{f}^\|_i,\\
\label{v-H-pha}
&\rho \ddot{v}_{i}^\bot+\D_{ij}\dot{v}_j^\bot-E_{ijkl}v_{k,lj}^\bot-D_{klij}v_{k,lj}^\|=
D_{klij}I^\|_{kl,j}+E_{ijkl}I^\bot_{kl,j}+\dot{f}^\bot_i.
\end{align}
The sources for the elastic displacement vectors are the gradient of the dislocation current tensors $\BI^\|$ and $\BI^\bot$, and the time derivative of the external forces $\Bf^\|$ and $\Bf^\bot$ . Eqs.~(\ref{B-H-pho})--(\ref{v-H-pha}) are hyperbolic
partial differential equations. Particularly, eqs.~(\ref{B-H-pho}) and (\ref{v-H-pho}), which describe the phonon fields, are partial differential equations of wave type, while
eqs.~(\ref{B-H-pha}) and (\ref{v-H-pha}), which describe the phason fields,
are partial differential equations of telegraph type. Equations of telegraph type \citep{Kneubuhl, MF} represent waves damped in time and propagating with finite velocity, so that causality and retardation are satisfied. The terms $\D_{ij}\dot{\beta}_{jm}^\bot$ and $\D_{ij}\dot{v}_j^\bot$ in eqs.~(\ref{B-H-pha}) and (\ref{v-H-pha}) respectively, are responsible for the damping and they appear due to the existence of the phason frictional forces. The tensor of characteristic time of damping  is defined as $(\tau_{ij})_{\text {tel}}=2 \rho \D_{ij}^{-1},$ where $\D_{ij}^{-1}$ is the inverse tensor of the $\D_{ij}$ \citep{AgiasofitouLazar2013}.

\item Static model: elastostatics of quasicrystals\\
In statics, all models conclude to the same equations for the elastic distortion fields. It is easy to see that
eqs.~(\ref{Eq127}) and (\ref{Eq129}), eqs.~(\ref{Eq130}) and (\ref{Eq131})
as well as eqs.~(\ref{B-H-pho}) and (\ref{B-H-pha})
give in the static limit the same equations for the elastic distortion  tensors $\Bbeta^\|$
and $\Bbeta^\bot$, in the absence of external forces
\begin{align}
\label{B-pho}
&C_{ijkl}\beta_{km,lj}^\|+D_{ijkl}\beta_{km,lj}^\bot=
e_{lmp}\large(C_{ijkl}{\alpha}^\|_{kp,j}+D_{ijkl}{\alpha}^\bot_{kp,j}\large),\\
\label{B-pha}
&E_{ijkl}\beta_{km,lj}^\bot+D_{klij}\beta_{km,lj}^\|=
e_{lmp}\large(E_{ijkl}{\alpha}^\bot_{kp,j}+D_{klij}{\alpha}^\|_{kp,j}\large).
\end{align}
Eqs.~(\ref{B-pho}) and (\ref{B-pha}) are coupled partial differential equations of Navier type.
The sources for the elastic distortion tensors are the gradient of
the dislocation density tensors $\Balpha^\|$ and $\Balpha^\bot$.
\end{itemize}
{\bf Note:} It should be noticed that in the case of vanishing phason fields, we can recover the dynamical dislocation model of classical elastodynamics (see, e.g., \cite{Kossecka77b}, \cite{Mura}) from all three models.

\section{The equations of motion for the displacement fields}
In dislocation dynamics it is also important to have the equations of motion for the displacement fields, in order to find the corresponding solutions. To this aim, we give the equations of motion for the displacement fields $\Bu^\|$ and $\Bu^\bot$ for the three considered models.
\begin{itemize}
\item Elastodynamic model of wave type:\\
By substituting the linear decompositions~(\ref{ugrad}) and (\ref{udot})
into the field equations~(\ref{EqM1}) and (\ref{EqM2}),
we obtain  the following inhomogeneous partial differential equations
\begin{align}
\label{u-pho-el}
&\rho \ddot{u}_i^\|-C_{ijkl}u_{k,lj}^\|-D_{ijkl}u_{k,lj}^\bot=
\rho \dot{v}_i^{\|\, \P}-C_{ijkl}{\beta}^{\|\, \P}_{kl,j}-D_{ijkl}{\beta}^{\bot\, \P}_{kl,j}+f_i^\|,\\
\label{u-pha-el}
&\rho \ddot{u}_i^\bot-E_{ijkl}u_{k,lj}^\bot-D_{klij}u_{k,lj}^\|=
\rho \dot{v}_i^{\bot\, \P}-D_{klij}{\beta}^{\|\, \P}_{kl,j}-E_{ijkl}{\beta}^{\bot\, \P}_{kl,j}+f_i^\bot,
\end{align}
where the plastic fields and the external forces are given as source terms.
Eqs.~(\ref{u-pho-el}) and (\ref{u-pha-el}) are partial differential equations of wave type, or in other words elastodynamic Navier equations. Both are hyperbolic partial differential equations. {\it Eqs. (\ref{u-pho-el}) and (\ref{u-pha-el}) are  the equations of motion for the displacement fields
in the theory of incompatible elastodynamics of wave type of quasicrystals}.

\item
Elasto-hydrodynamic model:\\
If we substitute the linear decompositions~(\ref{ugrad}) and (\ref{udot})
into the field equations~(\ref{EqM3}) and (\ref{EqM4}),
we obtain
\begin{align}
\label{u-pho-hy}
&\rho \ddot{u}_i^\|-C_{ijkl}u_{k,lj}^\|-D_{ijkl}u_{k,lj}^\bot=
\rho \dot{v}_i^{\|\, \P}-C_{ijkl}{\beta}^{\|\, \P}_{kl,j}-D_{ijkl}{\beta}^{\bot\, \P}_{kl,j}+f_i^\|,\\
\label{u-pha-hy}
&\D \dot{u}_i^\bot-E_{ijkl}u_{k,lj}^\bot-D_{klij}u_{k,lj}^\|=\D {v}_i^{\bot\, \P}-D_{klij}{\beta}^{\|\, \P}_{kl,j}-E_{ijkl}{\beta}^{\bot\, \P}_{kl,j}.
\end{align}
Eq.~(\ref{u-pho-hy}) is a wave type partial differential equation (elastodynamic Navier equation) for the phonon displacement field $\Bu^\|$, while eq.~(\ref{u-pha-hy}) is a diffusion type one for the phason displacement field $\Bu^\bot$. Eq.~(\ref{u-pho-hy}) is a hyperbolic partial differential equation, while eq.~(\ref{u-pha-hy}) is a parabolic one. Again the plastic fields and the external phonon body force are the source terms for the phonon and phason displacements.
{\it Eqs. (\ref{u-pho-hy}) and (\ref{u-pha-hy})
are  the equations of motion for the displacement fields
in the theory of incompatible elasto-hydrodynamics of quasicrystals}.

\item Elastodynamic model of wave-telegraph type:\\
Using eqs.~(\ref{ugrad}) and (\ref{udot}),
the field equations~(\ref{EqM5}) and (\ref{EqM6}) give a coupled system of partial differential equations
\begin{align}
\label{u-pho-AL}
&\rho \ddot{u}_i^\|-C_{ijkl}u_{k,lj}^\|-D_{ijkl}u_{k,lj}^\bot=
\rho \dot{v}_i^{\|\, \P}-C_{ijkl}{\beta}^{\|\, \P}_{kl,j}-D_{ijkl}{\beta}^{\bot\, \P}_{kl,j}+f_i^\|,\\
\label{u-pha-AL}
&\rho \ddot{u}_i^\bot+\D_{ij}\dot{u}_j^\bot-E_{ijkl}u_{k,lj}^\bot
-D_{klij}u_{k,lj}^\|=\rho \dot{v}_i^{\bot\, \P}
+\D_{ij}{v}_j^{\bot\, \P}
-D_{klij}{\beta}^{\|\, \P}_{kl,j}-E_{ijkl}{\beta}^{\bot\, \P}_{kl,j}+f_i^\bot,
\end{align}
where the plastic fields and the external phonon and phason body forces are
the source terms for the phonon and phason displacement vectors. Eq.~(\ref{u-pho-AL}) is a wave type partial differential equation (elastodynamic Navier equation) and
eq.~(\ref{u-pha-AL}) is a telegraph type partial differential equation. Both are hyperbolic partial differential equations. {\it Eqs.~(\ref{u-pho-AL}) and (\ref{u-pha-AL}) are  the equations of motion for the displacement fields
in the theory of incompatible elastodynamics of wave-telegraph type of quasicrystals}.

\item Static model: elastostatics of quasicrystals\\
In the static limit and in the absence of external forces, eqs.~(\ref{u-pho-el}) and (\ref{u-pha-el}), eqs.~(\ref{u-pho-hy}) and (\ref{u-pha-hy}), as well as eqs.~(\ref{u-pho-AL}) and (\ref{u-pha-AL}), reduce to
\begin{align}
&C_{ijkl}u_{k,lj}^\|+D_{ijkl}u_{k,lj}^\bot= C_{ijkl}{\beta}^{\|\, \P}_{kl,j}+D_{ijkl}{\beta}^{\bot\, \P}_{kl,j},\\
&D_{klij}u_{k,lj}^\|+E_{ijkl}u_{k,lj}^\bot=D_{klij}{\beta}^{\|\, \P}_{kl,j}+E_{ijkl}{\beta}^{\bot\, \P}_{kl,j}.
\end{align}
\end{itemize}
{\bf Remark:} We examine the case of compatible elastodynamics, which is important in studying problems of fracture mechanics, elastic waves, body forces, and in finding dispersion relations. In this case all the plastic fields are zero, and the above equations of motion (eqs.~(\ref{u-pho-el}) and (\ref{u-pha-el}), eqs.~(\ref{u-pho-hy}) and (\ref{u-pha-hy}), eqs.~(\ref{u-pho-AL}) and (\ref{u-pha-AL})) are simplified to the following equations.
\begin{itemize}

\item Elastodynamic model of wave type:
\begin{align}\label{comp1}
\rho \ddot{u}_i^\|-C_{ijkl}u_{k,lj}^\|-D_{ijkl}u_{k,lj}^\bot=f_i^\|, \quad
\rho \ddot{u}_i^\bot-E_{ijkl}u_{k,lj}^\bot-D_{klij}u_{k,lj}^\|=f_i^\bot.
\end{align}
Eqs.~(\ref{comp1}) are in agreement with the corresponding ones given by \citet{Ding1993}.

\item Elasto-hydrodynamic model:
\begin{align}
\label{}
\rho \ddot{u}_i^\|-C_{ijkl}u_{k,lj}^\|-D_{ijkl}u_{k,lj}^\bot=f_i^\|,\quad
\label{}
\D \dot{u}_i^\bot-E_{ijkl}u_{k,lj}^\bot-D_{klij}u_{k,lj}^\|=0.
\end{align}
The above equations have been derived for particular cases of quasicrystalline materials in the literature (see, e.g., \cite{Fan Book}). Applications to cracks can be found for instance in \citet{Zhu and Fan (2008), Wang et al. 2009}.

\item Elastodynamic model of wave-telegraph type:
\begin{align}
\label{}
\rho \ddot{u}_i^\|-C_{ijkl}u_{k,lj}^\|-D_{ijkl}u_{k,lj}^\bot=f_i^\|,\quad
\label{comp6}
\rho \ddot{u}_i^\bot+\D_{ij}\dot{u}_j^\bot-E_{ijkl}u_{k,lj}^\bot
-D_{klij}u_{k,lj}^\|=f_i^\bot.
\end{align}
\end{itemize}
Eqs.~(\ref{comp1})--(\ref{comp6}) have the same mathematical character with the corresponding ones (eqs.~(\ref{u-pho-el})--(\ref{u-pha-AL}))  of the incompatible case.

\section{Conclusions}
In this paper, we have investigated the fundamentals of the dynamics
of dislocations in quasicrystals.
We have examined the following models:
\begin{itemize}

\item
the elastodynamic model of wave type

\item
the elasto-hydrodynamic model

\item
the elastodynamic model of wave-telegraph type.
\end{itemize}
For these three models
we have derived the equations of motion of dislocations
in various forms suitable for different purposes. Through this presentation it has become evident, at least from the physical and mathematical point of view, that the elastodynamic model of wave-telegraph type offers a better physical representation for the dynamics of quasicrystals. Moreover, the derived equations provide the treatment of general dislocation motion
(with inertial effects) in quasicrystals. Like in classical elastodynamics, the source fields for the phonon and phason elastic distortion and the elastic
velocity fields are given in terms of phonon and phason dislocation density and dislocation current
tensors. These equations can build the basis for further research, such as to find
the solutions of the field equations for
uniformly and non-uniformly moving dislocations.
In addition, the results can be used for simulations
of discrete dislocation dynamics of quasicrystals
and also for experimental validation.

\section*{Acknowledgements}
The authors gratefully acknowledge grants from the
Deutsche Forschungsgemeinschaft (Grant Nos. La1974/2-1, La1974/3-1).

\end{document}